# Emission of coherent propagating magnons by insulator-based spin-orbit torque oscillators

M. Evelt[1], L. Soumah[2], A. B. Rinkevich[3], S. O. Demokritov[1,3], A. Anane[2], V. Cros[2], Jamal Ben Youssef[4], G. de Loubens[5], O. Klein[6], P. Bortolotti[2], V.E. Demidov[1*]

[1]*Institute for Applied Physics and Center for Nonlinear Science, University of Muenster, 48149 Muenster, Germany*

[2]*Unité Mixte de Physique, CNRS, Thales, Univ. Paris-Sud, Université Paris-Saclay, Palaiseau, France*

[3]*Institute of Metal Physics, Ural Division of RAS, Ekaterinburg 620108, Russia*

[4]*LABSTICC, UMR 6285 CNRS, Université de Bretagne Occidentale, 29238 Brest, France*

[5]*SPEC, CEA-Saclay, CNRS, Université Paris-Saclay, 91191 Gif-sur-Yvette, France*

[6]*SPINTEC, CEA-Grenoble, CNRS and Université Grenoble Alpes, 38054 Grenoble, France*

*Abstract:*

We experimentally demonstrate generation of coherent propagating magnons in ultra-thin magnetic-insulator films by spin-orbit torque induced by dc electric current. We show that this challenging task can be accomplished by utilizing magnetic-insulator films with large perpendicular magnetic anisotropy. We demonstrate simple and flexible spin-orbit torque devices, which can be used as highly efficient nanoscale sources of coherent propagating magnons for insulator-based spintronic applications.

*Corresponding author. e-mail: demidov@uni-muenster.de

Recent advances in the studies of the spin-orbit torque [1-4] opened novel perspectives for nano-spintronics by enabling utilization in spintronic devices of insulating magnetic materials such as Yttrium Iron Garnet (YIG), which is famous for its unprecedented small magnetic damping [5]. The small natural damping is crucial for the reduction of power consumption of spintronic devices based on the spin-transfer torque phenomena [6,7]. Additionally, the small damping is highly beneficial to applications utilizing propagating magnons as carriers of spin information, since it results in large magnon propagation length [8-11].

It is now well established that spin-orbit torque (SOT) in YIG/Pt systems can be used to enhance the propagation of coherent magnons excited by microwave-frequency field [12-14] and to achieve the complete compensation of the natural damping resulting in the spontaneous excitation of coherent magnetization auto-oscillations [15-17]. It is also experimentally and theoretically ascertained, that, in the regime of small driving currents below the damping compensation point, SOT causes excitation of incoherent magnons with a broad spectral distribution [18,19]. These magnons are able to propagate in YIG films to significant distances [9, 20-23] and can be used for controllable transmission of spin information [24]. However, the lack of coherence does not allow one to utilize these magnons for information processing based on coherent-wave phenomena, where the phase of the waves plays a decisive role [25].

Transition from the incoherent to the coherent regime is expected to occur when SOT completely compensates the natural damping in the magnetic film resulting in the excitation of coherent auto-oscillations [15,16], which can be treated as an accumulation of a large number of the SOT-excited magnons in a single spectral state [19]. However, this process is systematically observed in experiments to be accompanied by a nonlinear shift of the magnon frequency, which causes a spatial self-localization of coherent magnons preventing their emission into the surrounding magnetic film [15,16,23,26].

Here we experimentally demonstrate an approach, which allows one to overcome the self-localization phenomena and achieve an efficient emission of SOT-excited coherent magnons into an extended magnetic-insulator film. We show that this can be accomplished by utilizing nanometer-thick Bi-substituted YIG films exhibiting large perpendicular magnetic anisotropy (PMA) [27]. The effects of PMA result in a compensation of the in-plane shape anisotropy of the film. Due to this compensation, the nonlinear frequency shift vanishes leading to the current-independent frequency of the spontaneously formed coherent magnon state. As the result, the created state can emit propagating magnons with the wavelength of about 300 nm. We demonstrate that the Bi doping necessary to achieve sufficiently strong PMA has no significant adverse effect on the density of the electrical current necessary to excite SOT-induced auto-oscillations and on the magnon propagation length in YIG. This makes the suggested approach very promising for insulator-based spintronic and magnonic applications.

The schematic of our experiment is shown in Fig. 1. The studied devices are based on an extended 20-nm thick film of Bi-doped YIG (BiYIG) ($Bi_1Y_2Fe_5O_{12}$) with PMA grown by the pulsed laser deposition on substituted Gallium Gadolinium Garnet (sGGG) substrate [27] (see Supplementary information for film characterization). A 6-nm thick Pt film is sputter-deposited onto BiYIG and is patterned into a shape of a strip line with the width of 1 μm and the length of 4 μm. Due to the spin-orbit interaction in Pt [1,2] causing the scattering of conduction electrons with opposite magnetic moments toward opposite surfaces of the Pt film, the dc electric current $I$ flowing in the plane of the strip line [28] is converted into the out-of-plane spin current $I_s$ (see inset in Fig. 1).

The spin current is injected into the BiYIG film and exerts a torque on its magnetization *M*. The saturating static magnetic field $H_0$ is applied in the device plane and perpendicular to the direction of the current flow. For the orientation of $H_0$, as shown in Fig. 1, the positive current *I* results in SOT compensating the natural damping in the BiYIG film [29]. When the damping is completely compensated, large-amplitude coherent magnetization auto-oscillations are excited in the magnetic film above the Pt line. The auto-oscillations can emit coherent magnons into the surrounding BiYIG film, provided that the frequency of the auto-oscillations is not shifted below the spectrum of propagating magnons due to the nonlinear frequency shift.

We detect the SOT-induced magnetization oscillations and the emitted magnons by the space- and time-resolved micro-focus Brillouin light scattering (BLS) spectroscopy [30]. We focus the probing laser light with the wavelength of 532 nm and the power of 0.1 mW into a diffraction-limited spot on the surface of the BiYIG film through the sample substrate (Fig. 1) and analyze the spectrum of inelastically scattered light. The resulting BLS signal is proportional to the density of magnons or, alternatively, to the intensity of magnetization oscillations at the position of the probing spot. By rastering the probing-light spot over the surface of the sample, the spatially resolved maps of the magnon density can be obtained.

In Figure 2(a), we present the color-coded BLS intensity in the current-frequency coordinates measured by placing the probing spot in the middle of the Pt line. We find that, at *I*≥1.8 mA, the BLS spectra exhibit a narrow intense peak (see also inset in Fig. 2(a)) corresponding to the SOT-induced auto-oscillations in the BiYIG film. The onset current $I_C$=1.8 mA corresponds to the current density in the Pt layer of $3\times10^{11}$ A/m$^2$, which is close to the value obtained for devices based on undoped in-plane magnetized

YIG films [15]. In contrast to the previously demonstrated devices, where the frequency of auto-oscillations strongly decreased with the increase of $I$ due to the nonlinear frequency shift [15,26], in BiYIG-based devices, the frequency of the oscillations stays nearly constant over the entire range of $I$, while the intensity of auto-oscillations monotonously increases with the increase of the driving current (Fig. 2(b)). The frequency of the auto-oscillations $f_A$=5.56 GHz is close to the independently determined frequency of the ferromagnetic resonance (FMR) in the BiYIG/Pt bilayer $f_0$=5.63 GHz (see Supplementary information). Therefore, one can characterize the nonlinear frequency shift by utilizing the Kittel formula for the FMR frequency in an in-plane saturated magnetic film with PMA [31]

$$f_0 = \gamma\sqrt{H_0(H_0 + (4\pi - N^a)M_e)} \ . \qquad (1)$$

Here $\gamma$ is the gyromagnetic ratio, $N^a = H_a/M_e$ is the magnetization-independent demagnetization factor associated with the anisotropy, $H_a$ is the effective anisotropy field, and $M_e$ is the effective static magnetization, which decreases with the increase in the current in the Pt layer resulting in the variation of the frequency $f_0$. For BiYIG films used in our study $N^a \approx 4\pi$. Therefore, the $M_e$-dependent term in Eq. (1) vanishes resulting in the vanishing nonlinear frequency shift.

Since the nonlinear frequency shift is known to be responsible for the localization of coherent auto-oscillations, one expects that, in the studied devices, the fulfilment of the magnon propagation conditions will be restored leading to emission of coherent magnons by auto-oscillations excited in the active device area. Indeed, spatially resolved BLS measurements (Fig. 3(a)) confirm the efficient magnon emission. Figure 3(a) shows the color-coded spatial map of the intensity of SOT-induced magnetization oscillations recorded by rastering the probing laser spot over 20 μm by 7 μm area centered at the Pt line (dashed lines in Fig. 3(a)). As seen from these data, the oscillations excited due to SOT are not localized in the active area above the Pt line, but significantly extend into the surrounding BiYIG film suggesting the emission of excited coherent magnons. Figure 3(b) shows the $x$-dependence of the integrated BLS intensity on the log-linear scale. The dependence exhibits a well-defined exponential decrease enabling determination of the mean propagation length of the emitted magnons $\xi$=3.7 μm. The spatially resolved measurements also allow us to determine the group velocity of the magnons by comparing the temporal dependences of the BLS intensity at $x$=0 and 10 μm (Fig. 3(c)). As seen from these data, the pulse of the dynamic magnetization detected at $x$=10 μm is delayed with respect to that at $x$=0 by about 58 ns, which corresponds to the group velocity $V_g$≈0.17 μm/ns.

To get insight into the nature of excited magnons, we calculate the magnon dispersion spectrum in the BiYIG film using the analytical theory developed in Ref. [31] and independently determined anisotropy field $H_a$=1.84 kOe (see Supplementary information). Figure 4(a) shows the calculated dispersion curves for magnons propagating parallel ($\phi$=0) and perpendicular ($\phi$=90°) with respect to the direction of the static magnetic field. The dashed horizontal line marks the experimentally determined frequency of the auto-oscillations $f_A$ (Fig. 2(a)). As seen from Fig. 4(a), $f_A$ is located within the spectrum of propagating magnons allowing their emission from the active device area. We emphasize that magnons with the frequency $f_A$ can propagate in all directions with respect to the static field (see the isofrequency curve in the inset in Fig. 4(a)), which is in agreement with the experimental emission pattern (Fig. 3(a)). Note that, as shown by Fig. 4(b), magnons emitted at different angles $\phi$ are expected to possess significantly different wavelengths and group velocities. In particular, magnons emitted in the direction perpendicular to the Pt line ($\phi=0$) exhibit the shortest wavelength of about 340 nm and the largest group velocity of 0.21 µm/ns. The latter value is reasonably close to that obtained in the experiment, which confirms the validity of our calculations.

Finally, we discuss the dependence of the characteristics of emitted magnons on the magnitude of the static magnetic field $H_0$. Figure 5(a) shows the field dependence of the experimentally determined magnon propagation length $\xi$ and its inverse value $k''=1/\xi$ – the imaginary part of the magnon wavevector. As seen from these data, the propagation length significantly increases at small $H_0$, which is associated with the reduction of the magnon relaxation frequency: $\omega_r = \omega_{r0} + \alpha\gamma H_0$, where $\alpha$ is the Gilbert damping parameter and $\omega_{r0}$ is a nonzero offset originating from the influence of the inhomogeneity of the magnetic film. In agreement with the theory [32], $k''=\dfrac{\omega_{r0}}{V_g}+\alpha\dfrac{\gamma H_0}{V_g}$ varies linearly with $H_0$ (line in Fig. 5(a)). Taking into account the experimentally determined group velocity $V_g$=0.17 µm/ns, which was found to be nearly constant within the range $H_0$=1-2 kOe, we calculate from the slope of the experimental dependence $k''(H_0)$ the damping parameter for the BiYIG film $\alpha\approx8.2\times10^{-4}$. This value is not much larger than that typical for undoped YIG films of the same thickness [15, 33]. However, it is significantly smaller in comparison with the values typical for metallic magnetic films, making BiYIG a promising material for insulator-based spintronic applications. Note that the found value characterizes the damping in a free BiYIG film, while, in a BiYIG/Pt bilayer, the damping is expected to be significantly increased due to the influence of the spin pumping effect. To characterize the damping in the active device area, we analyze the field dependence of the auto-oscillation onset current (Fig. 5(b)), which is proportional to the relaxation frequency [34]: $I_C = \beta(\omega_{r0} + \alpha_{Pt/YIG}\gamma H_0)$, where $\beta$ is the proportionality coefficient describing the SOT efficiency. Using the results of the previous analysis, we find the Gilbert damping constant $\alpha_{Pt/YIG} \approx 1.6 \times 10^{-3}$. Since in a bilayer constituted by a nanometer-thick low-damping insulator and Pt, the damping is mostly determined by the spin pumping, the above value is very close to that found for bilayers based on undoped YIG [15]. This fact elucidates the similarity in the auto-oscillation onset current density between the devices studied in our work and those based on undoped YIG films.

In conclusion, we have experimentally demonstrated devices capable of emission of coherent spin-orbit torque excited magnons into magnetic-insulator films. We show that this can be achieved by using films with perpendicular magnetic anisotropy, where the effects of the anisotropy allow one to overcome the nonlinear localization of magnons. The demonstrated devices are flexible with respect to modifications and can be used as coherent sources for magnon-based nanoscale information transmission and processing. We believe that our results will stimulate further developments in the emerging field of insulator-based spintronics.

**ACKNOWLEDGMENTS**

We thank C. Carrétéro and E. Jacquet for their contribution to films growth. We acknowledge support from Deutsche Forschungsgemeinschaft, FASO of Russia (theme "Spin" No. AAAA-A18-118020290104-2), Russian ministry of Education and Science (project No. 14.Z50.31.0025), and that of the ANR Grant ISOLYIG (ref 15-CE08-0030-01). LS is partially supported by G.I.E III-V Lab. France.

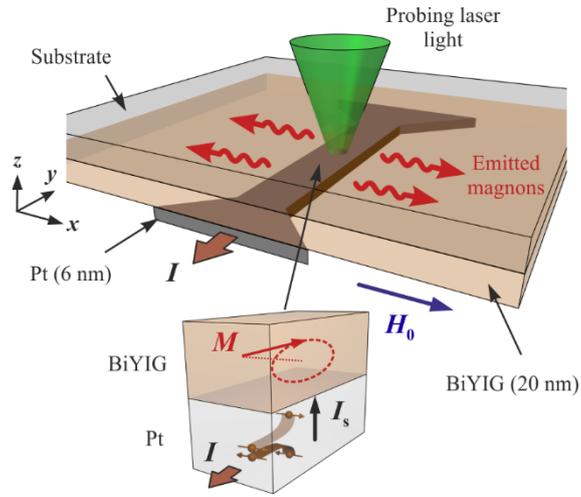

**Figure 1.** Experimental layout. The device consists of an extended 20-nm thick BiYIG film with PMA and 6-nm thick Pt line with the length of 4 µm and the width of 1 µm. The dc current $I$ in Pt is converted into the pure spin current $I_s$ (see the inset) inducing local coherent magnetization auto-oscillations, which emit propagating coherent magnons into the surrounding BiYIG film.

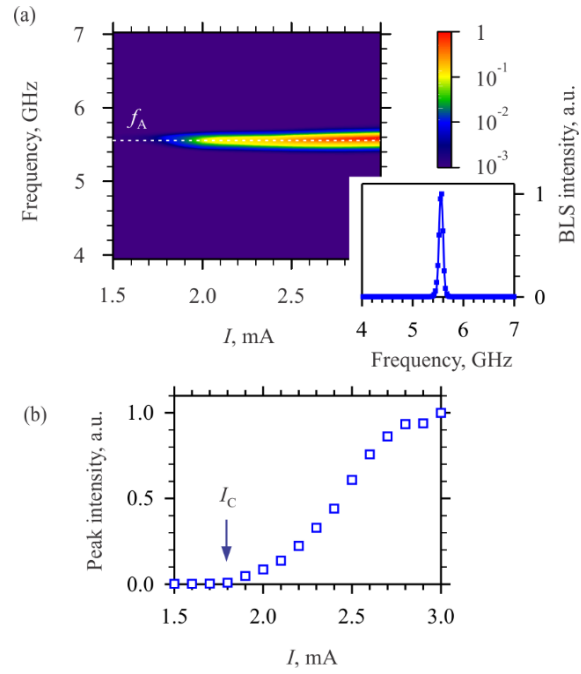

**Figure 2.** (a) Color-coded BLS intensity in the current-frequency coordinates measured by placing the probing spot in the middle of the Pt line. $f_A$ marks the frequency of SOT-induced auto-oscillations. Inset shows the representative BLS spectrum for $I$=2.5 mA. (b) Current dependence of the peak BLS intensity. $I_C$ marks the current corresponding to the onset of auto-oscillations. The data were obtained at $H_0$=2.0 kOe.

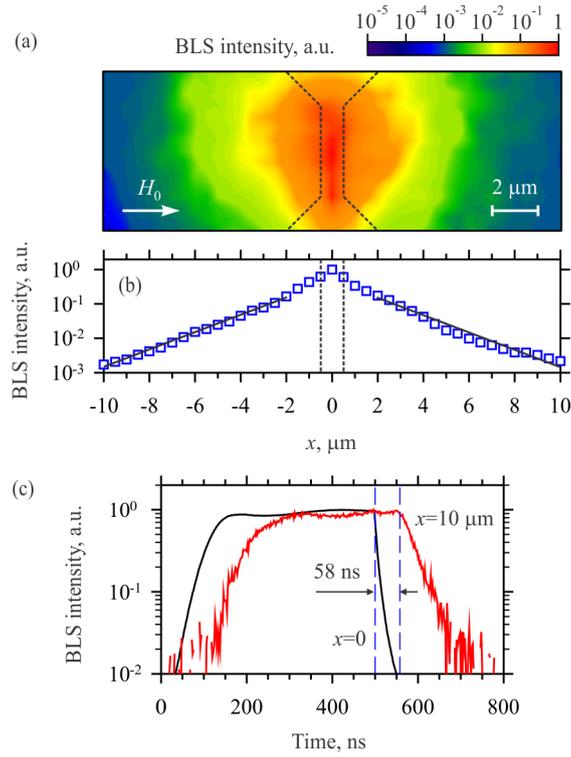

**Figure 3.** (a) Color-coded spatial map of the BLS intensity recorded by rastering the probing laser spot over 20 μm by 7 μm area. Dashed lines show the contours of the Pt line. (b) *x*-coordinate dependence of the integrated BLS intensity on the log-linear scale. Symbols – experimental data, solid curves – exponential fit at |*x*|>2 μm. Vertical dashed lines show the boundaries of the Pt line. (c) Temporal dependences of the BLS intensity recorded at *x*=0 and 10 μm. Dashed vertical lines mark the beginning of the rear edge of the pulses used to determine the magnon propagation delay. The data were obtained at $H_0$=2.0 kOe and *I*=2.5 mA.

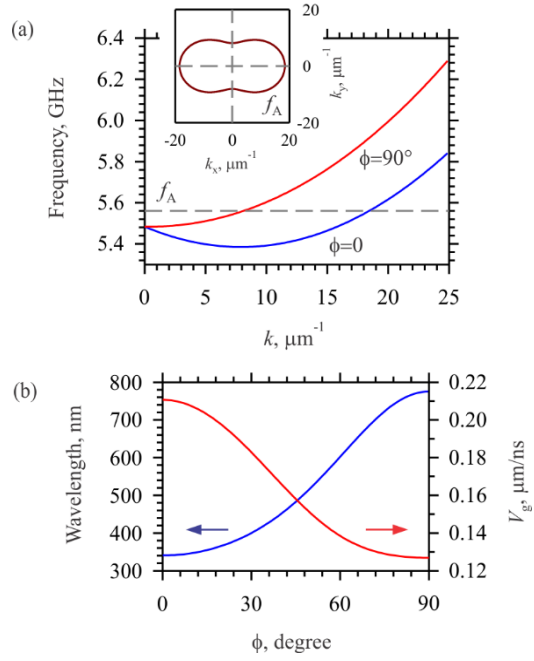

**Figure 4.** (a) Calculated dispersion curves for magnons propagating parallel ($\phi=0$) and perpendicular ($\phi=90°$) with respect to the direction of the static magnetic field. The dashed horizontal line marks the experimentally determined frequency of the auto-oscillations $f_A$. Inset shows the isofrequency curve at the frequency $f_A$ in the $k_x$-$k_y$ coordinates, where $k_x$ and $k_y$ are the $x$- and $y$-components of the total wavevector $k$. (b) Calculated angular dependences of the wavelength and the group velocity of magnons at the frequency $f_A$. The calculations were performed for $H_0=2.0$ kOe using the independently determined effective field of PMA $H_a=1.84$ kOe.

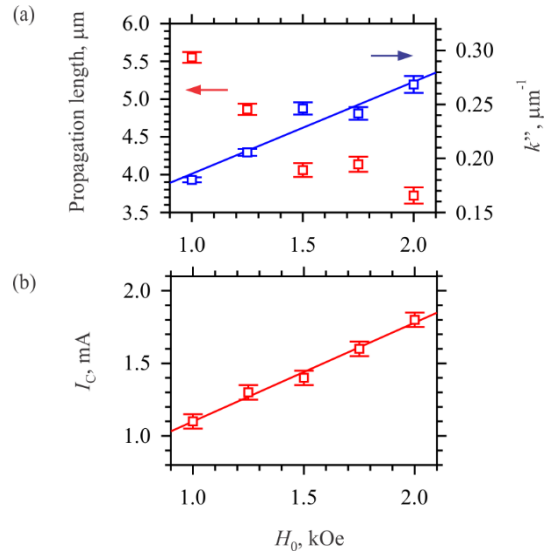

**Figure 5.** (a) Field dependence of the experimentally determined magnon propagation length and of the imaginary part of the magnon wavevector $k''$. Symbols – experimental data. Solid line – linear fit of the data for $k''$. The data were obtained at $I$=2.5 mA. (b) Field dependence of the auto-oscillation onset current. Symbols – experimental data. Solid line – linear fit.